\documentclass[twocolumn]{elsart}

\usepackage{amssymb}
\usepackage{epsfig}
\usepackage{cite}
\newlength{\bildtitel}
\renewcommand{\baselinestretch}{1}

\begin{document}

\begin{frontmatter}



\title{A microfabricated sensor for thin dielectric layers}

\renewcommand{\baselinestretch}{1}

\author{P.~Fierlinger, R.~DeVoe, B.~Flatt, G.~Gratta, M.~Green,}
\author{S.~Kolkowitz, F.~Leport, M.~Montero Diez, R.~Neilson,}
\author{K.~O'Sullivan, A.~Pocar, J.~Wodin$^1$}
\address{Physics Department, Stanford University, Stanford, CA 94305} 
\begin{abstract}
We describe a sensor for the measurement of thin dielectric layers capable of operation
in a variety of environments.  The sensor is obtained by microfabricating a capacitor 
with interleaved aluminum fingers, exposed to the dielectric to be measured.   
In particular, the device can measure 
thin layers of solid frozen from a liquid or gaseous medium.    Sensitivity to single
atomic layers is achievable in many configurations and, by utilizing fast, high sensitivity 
capacitance read out in a feedback system onto environmental parameters, coatings of few
layers can be dynamically maintained.    We discuss the design, read out and calibration
of several versions of the device optimized in different ways. We specifically dwell on the 
case in which atomically thin solid xenon layers are grown and stabilized, in cryogenic conditions,
from a liquid xenon bath.
\end{abstract}
\end{frontmatter}
\vfill
\small{$^1$Now at 
Stanford Linear Accelerator Center, Menlo Park, CA 94025}
\renewcommand{\baselinestretch}{1}
\section{Introduction}
The ability to detect atomically thin layers of materials deposited onto a substrate is
essential in many branches of science and technology.    By far the most common method
employs ``microbalances''~\cite{microbalances}, piezoelectric quartz oscillators whose
resonant frequency changes as the material deposited onto the crystal increases the
mass associated with its surface atomic layer.    Microbalances, however, can only produce
high resolution measurements if they operate in a medium that does not substantially
reduce their mechanical quality factor.   In particular they are not suitable for operation
in liquids that, because of viscosity, substantially damp the crystal oscillations.

The sensor development discussed here arises from the need to measure thin layers of
solid xenon (SXe) grown on a probe that is immersed in liquid xenon (LXe) at $\sim$ 170~K
and further cooled.   In the context of the EXO neutrino-less double-beta ($0\nu\beta\beta$)
detector R\&D program~\cite{exo}, the need arises to retrieve and identify single
barium ions produced by the $\beta\beta$ decays of $^{136}$Xe.    In a possible design of the
experiment, several tons of $^{136}$Xe in liquid phase would serve as the source of 
$0\nu\beta\beta$ decay as well as the detection medium in a time projection chamber
(``TPC''~\cite{TPC}).   Candidate decays would be localized by the TPC and a mechanical
extraction device would be inserted in the LXe, retrieve the Ba-ion and transport it into
an ion trap.    While single Ba-ion identification in a trap filled with buffer gases
including Xe has already been achieved~\cite{iontrap,jessePRL}, the device to transport
single ions from LXe to the low pressure ion trap with high efficiency is still being
developed.     One technology under investigation consists of electrostatically attracting
the Ba-ion onto a probe tip coated with SXe. Due to its large ionization potential, 
SXe isolates the chemically active ion 
from the substrate material and makes subsequent release possible, by sublimation of the 
coating in vacuum.    The requirement
that only a very small amount of Xe be released in the ion trap calls for the development of
a SXe sensor capable of operation in LXe with sensitivity of a few atomic layers. 

The sensor is based on a microfabricated capacitor composed of two arrays of
interleaved metallic fingers on a dielectric substrate.   The fingers are exposed to the
environment to be measured so that their mutual capacitance depends on the dielectric
constant $\epsilon_r$ of the environment, assumed to be dielectric.   The possibility
of making extremely sensitive capacitance measurements allows for detecting 
very small changes in $\epsilon_r$. While only capacitance measurements
are used in the rest of this work, the combined measurement of capacitance
and loss factor as function of the frequency is a common way of characterizing bulk
dielectrics~\cite{ref:dielectrics} and could extend the range of application for devices
of the type discussed here.  Possible applications include the measurement of thin layer 
deposition in gaseous and liquid environments, the detection of certain chemical species in gas and 
liquids or the fast analysis of certain immiscible (multiphase) liquids (e.g.\,oil-water emulsions). 
The clean, rugged, simple and small design of the device is suitable for a variety of 
environments, from UHV to in-situ geological measurements.

\section{Description and fabrication of the device}\label{sec:descr}

\begin{figure}[!tb]
  \begin{center}
 \includegraphics[scale = 0.19]{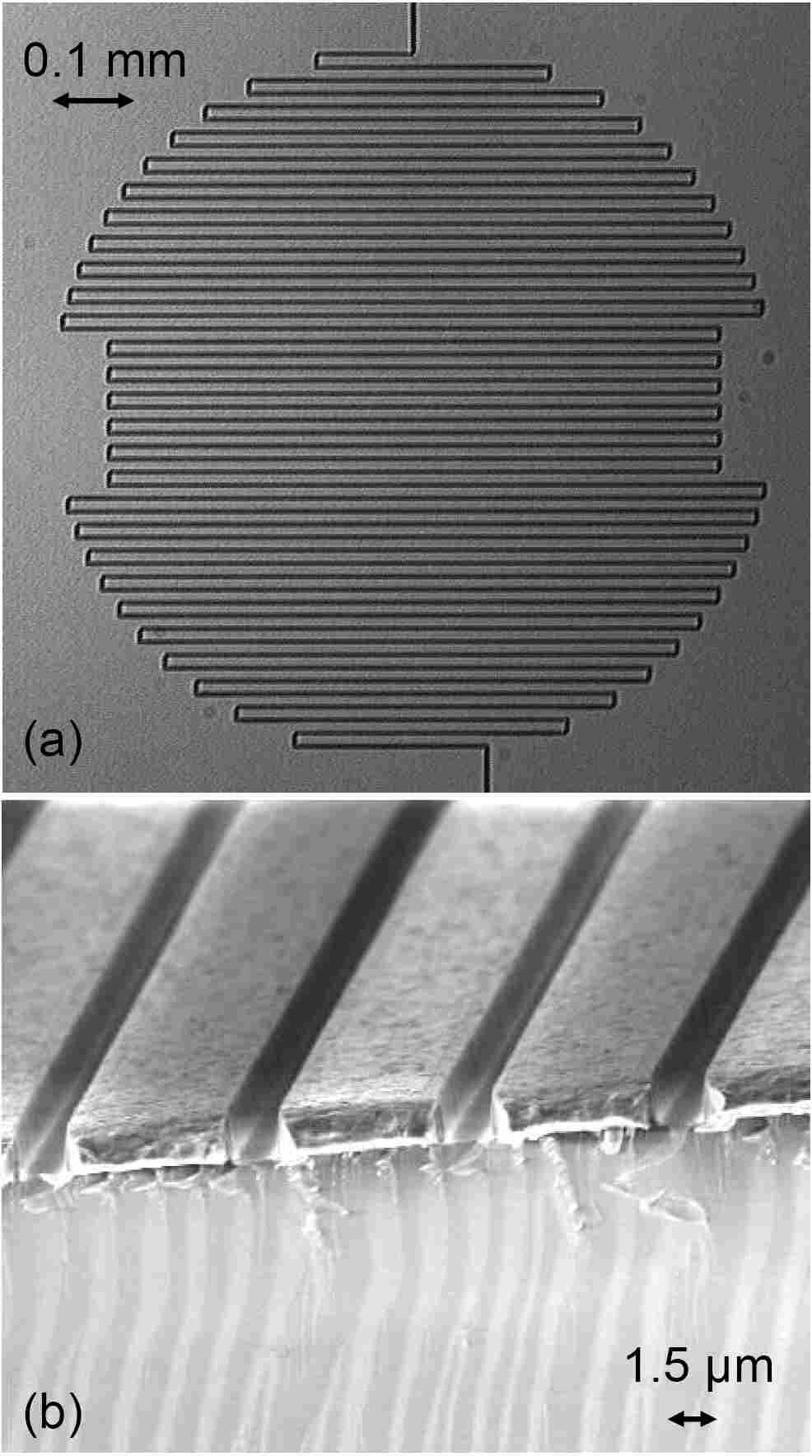}
  \caption{Details of a 0.94~mm diameter sensor.    Panel (a) shows a
sensor with 5.5~$\mu$m aluminum strip width and 1.5~$\mu$m spacing.   The pads 
for the external connections are clearly visible on the left and right side of the image.     
Panel (b) shows an electron
microscope photograph of the broken edge of a sensor.    The aluminum strips are
1.85~$\mu$m-thick.    The frozen layer to be measured grows on the structure and
fills the ``canyons'' between strips to different degrees, altering the capacitance.
}
\label{fig:sensorpic}
\end{center}
 \end{figure}

A magnified picture of a sensor is shown in Fig.~\ref{fig:sensorpic} (a). 
The darker line represents the gap between the aluminum fingers.  The overall circular configuration
makes the sensor suitable for mounting at the end of a small-diameter ($\sim$~1-2~mm)
probe and is otherwise irrelevant for the performance of the device.   The test device
shown in Fig.~\ref{fig:sensorpic} (a) is not yet cut in the final circular shape. 
A 200~$\mu$m-thick quartz wafer is used as substrate.   Various structures with 
different Al-strip width $w$, spacing $s$, thickness $t$ and
overall diameter $D$ are produced, resulting in capacitances up to 30~pF in vacuum. 
All devices discussed here are produced using optical lithography.    

Different parameters are optimized as follows: the overall diameter $D$ has to be
small for our application, so that the device can be mounted on the tip of a
thin probe capable of being inserted in the LXe detector as well as into an ion
trap.    Devices were produced with $D$ from 0.94 to 1.9~mm. 
The spacing between the metallic fingers and their thickness are important to obtain
good sensitivity for a particular range of thicknesses of frozen material.  
We chose $t\simeq 1.8$~$\mu$m and $s\simeq 1.5$~$\mu$m, so that the spaces to be filled
by the medium under measurement have a roughly square cross section.   The width
of the aluminum strips was varied quite substantially in different devices as
thin strips provide larger capacitance per unit area but may be more fragile
(unlike most electronic devices, this sensor has to be exposed directly to
different fluids).
The devices discussed here have capacitances in vacuum in the range of 2~pF to 30~pF.
The series resistance of the devices (due to the small cross section of the strips)
is in all cases negligible with respect to the reactance at the frequencies
(1~kHz to 1~MHz) used for the capacitive measurements.

Device fabrication begins with coating the top side of a quartz wafer with 2~$\mu$m
aluminum.   The Al is then covered with 1.6~$\mu$m Shipley 3612 positive
photoresist that is then exposed to 350~nm light for 1.6~s through a vacuum
contacted photolithography mask.  After developing, the structure is etched into
the Al layer using reactive ion plasma etching with optical endpoint detection
in an Applied Materials P5000 etcher. The residual photoresist is removed
using SVC PRX-127 resist stripper. The line spacing of $s = 1.5$~$\mu$m
is close to the technical limit for using contact masks. A scanning electron microscope 
picture of the walls of the etched channels is shown in Fig.~\ref{fig:sensorpic} (b).
Wafers are covered with protective tape, diced in a square shape and directly used or 
later ground into a roughly circular shape.    The yield of this rather crude 
process is 50\%, sufficient for our purpose, although clean circular cutting could 
be achieved using standard techniques (e.g. bead blasting or laser cutting).

An overview of all sensors produced is given in Tab.~\ref{tab:sensors}.
The capacitance values in this table are measured in air and hence are
intended only for generic sensor characterization.    The measurements
for the table are obtained at a readout frequency of 10~kHz, but no significant change in capacitance
is observed in the 100~Hz - 2~MHz frequency range explored.    Taking 
edge effects into account, the capacitance scales with the geometrical dimensions 
as expected. It should be noted that we also produced sensors with $w=300$~nm and $s\sim100$~nm 
using electron-beam lithography and plasma etching of 300~nm thick aluminum. These structures were found  
to be too small to grow consistent frozen layers between strips and will not be 
discussed further.

\section{Cryogenic setup and readout}\label{sec:readout}

The sensor can be configured in many different ways to detect thin layers of  
different dielectric materials of interest.     While testing with water 
and other materials freezing near room temperature is rather straight forward, 
here we describe a more elaborate system that was developed to make tests at 
cryogenic conditions and, in particular, maintain stable SXe layers at a variety of 
pressures by cooling the sensor with liquid helium (LHe) to arbitrary temperatures 
as low as $\simeq$~20~K. The setup also allows to heat the sensor up to $\simeq$~650~K 
to sublimate the coating. 
\begin{figure}[!tb]
  \begin{center}
 \includegraphics[scale = 0.075]{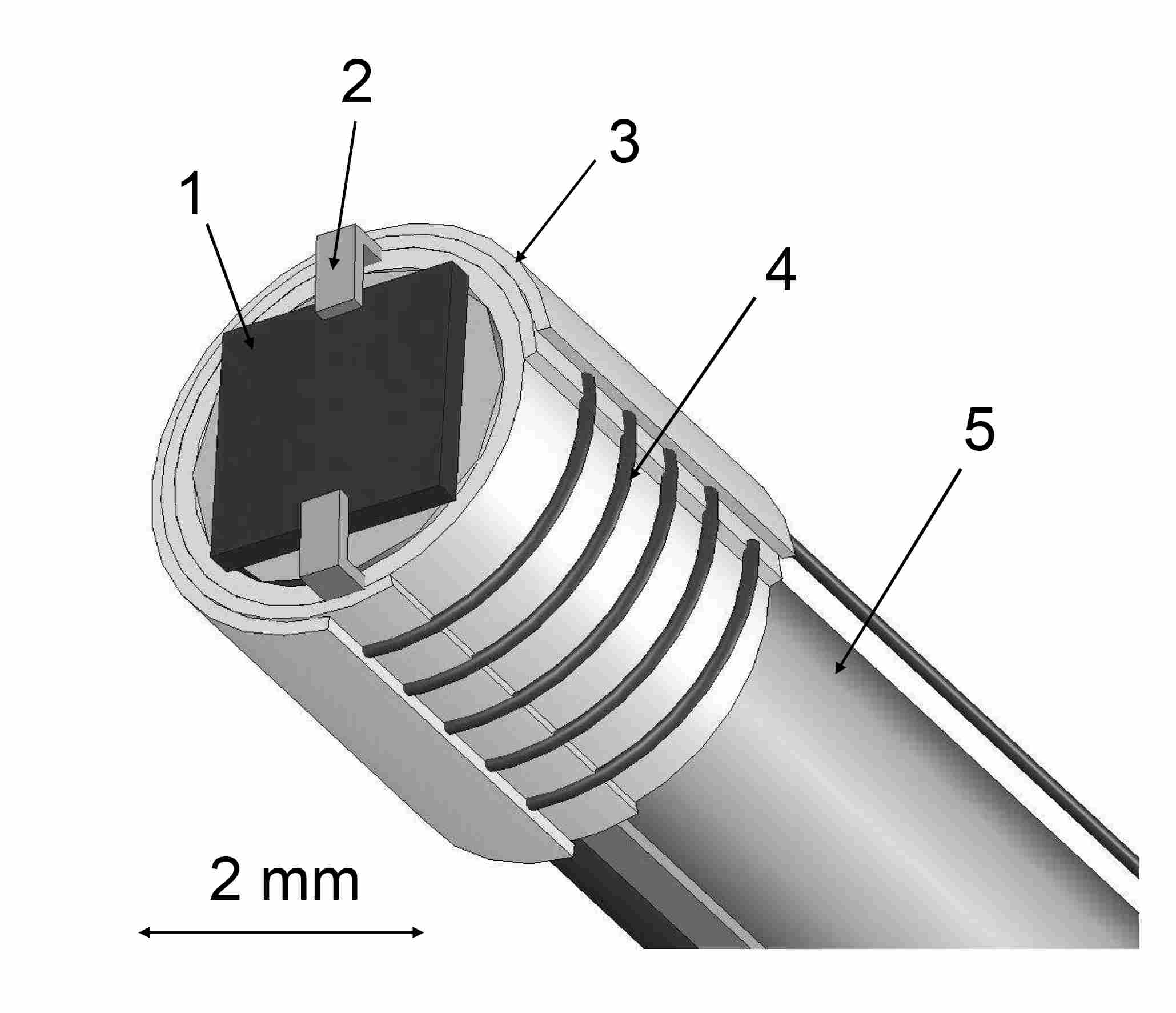}
  \caption{Diced 2~mm diameter sensor (1) mounted at the end of the copper cold finger (5) 
  of the miniaturized cryostat. The readout pads of the sensor are contacted by gold 
  coated leads (2). An assembly of three Vespel sleeves (3) holds the leads and the 
tungsten heater wire (4) in place.}
\label{fig:sensor_on_tip}
\end{center}
 \end{figure}
Most of the data presented in this 
paper was collected with this advanced setup or with simpler arrangements. 
The sensor is thermally and mechanically connected to a copper cold finger using  
VGE-7031 varnish or indium. Gold-plated readout leads are pressed against the sensor's 
Al pads and a thin walled Vespel\cite{vespel} polyimide tube section is forced around the leads 
to make the setup sturdy and strain relief the contacts.   UHV-rated 
ultraminiature polyimide-insulated coaxial cables type "SC" 
from Lakeshore\cite{lakeshore} are run along the stem of the cryostat and bring the signals to room temperature.  
Figure~\ref{fig:sensor_on_tip} schematically shows one device mounted on the tip 
of the cold finger, together with a 0.05~mm tungsten heater wire.

\begin{figure}[!htb]
  \begin{center}
 \includegraphics[scale = 0.2]{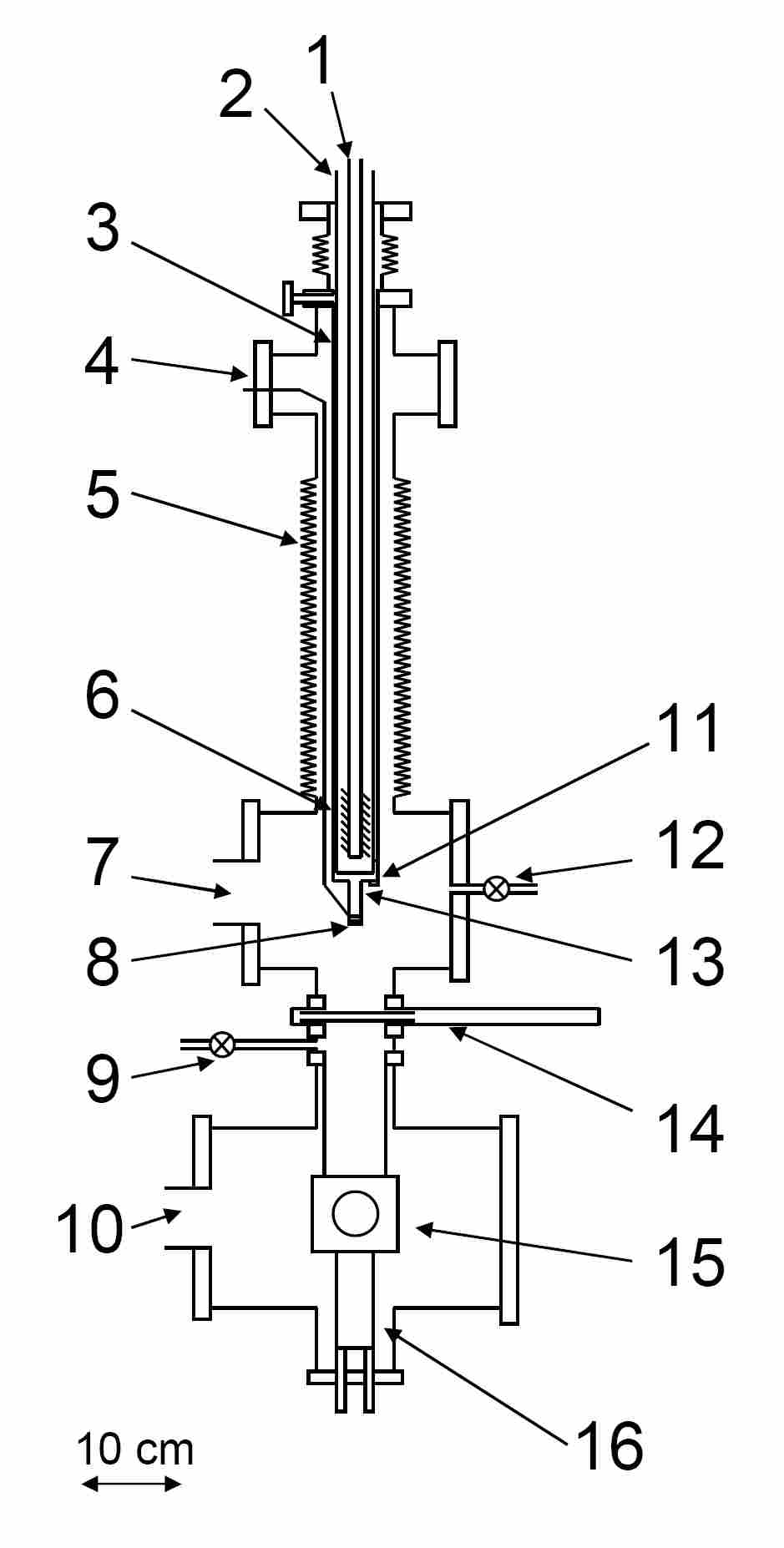} 
  \caption{Schematic drawing of the cryogenic setup allowing to grow frozen  
layers from liquid or gas phase  
for a variety of materials.  (1) insulated LHe supply line, (2) He return line,  
(3) cryostat insulation vacuum pumpout manifold, (4) electrical feed through,  
(5) vertical motion bellows, (6) heat exchanger, (7) medium pumping port,  
(8) sensor and heater assembly, (9, 12) gas manifold, (10) liquid Xe cell insulation vacuum  
pumpout manifold, (11) Si diode temperature sensor, (13) Cu cold finger, (14) gate valve,  
(15) liquid Xe cell, (16) heat exchanger assembly with LN2 lines, heater and temperature sensor. 
The cold finger and the sensor/heater assembly are showed enlarged in Fig.~\ref{fig:sensor_on_tip}.} 
\label{fig:cryotip}
\end{center}
 \end{figure}

\begin{figure}[tb]
  \begin{center}
 \includegraphics[scale = 0.14]{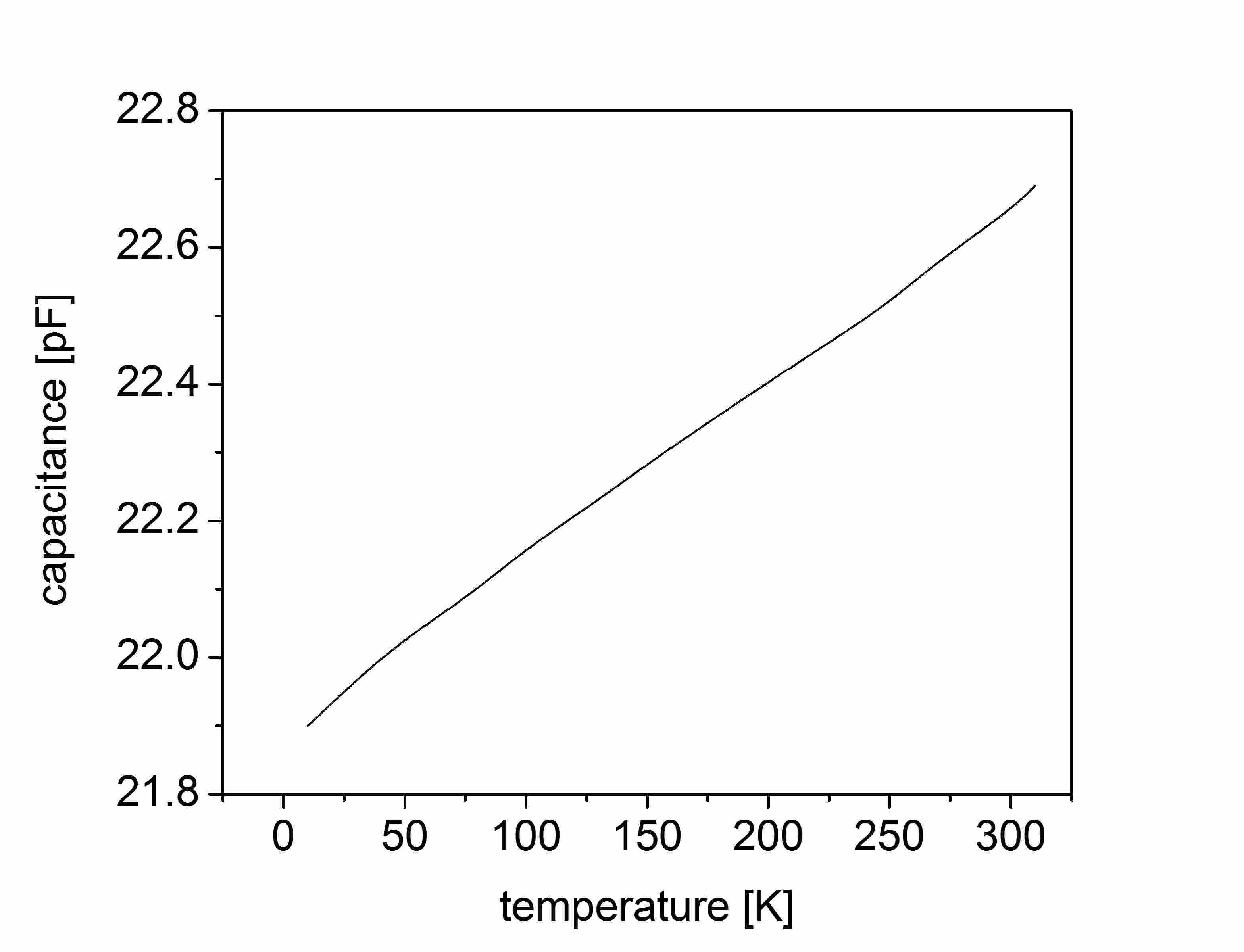}
\caption{Capacitance of a sensor during cooling from 
room temperature to about 15\,K in 2$\cdot$10$^{-8}$\,mbar (300~K) vacuum. 
The capacitance is the  
sum of the unshielded sections of the readout cables (0.7~pF) with the dielectric constant of 
vacuum. The fluctuations of the measurement are smaller than 
the line width in the plot.}
\label{vac}
\end{center}
\end{figure}

The cold finger is mounted in the setup schematically shown in 
Fig.~\ref{fig:cryotip}.    With reference to the figure, LHe is injected through 
the central vacuum insulated line from the top of the cryostat (1),  
it cools the cold finger (13) and sensor (8) through a copper heat exchanger (6)  
and is exhausted from the coaxial line (2).      
The entire LHe assembly is surrounded by a 9~mm diameter vacuum jacket (3). 
The tungsten heater wrapped around the cold finger has a maximum power of about 2.5~W and 
is used to fine tune the temperature of the tip and selectively evaporate frozen layers.    
The cooled probe can be translated vertically by 40~cm with a set 
of bellows (5) and a pneumatic actuator (not shown), so that it can be inserted into a small 
LXe cell (15) cooled by a second, larger cold finger (16) submerged in liquid nitrogen (LN$_2$).   
With this setup, layers can be grown from gas or liquid phase from a variety of media. 
All temperatures and pressures are readout by a LabView\cite{labview} control system 
that operates valves, heaters and the vertical position of the probe, allowing for fully 
automated operation of the system.    Because of the small dimensions of the 
probe assembly only two coaxial wires are used for the connection of the sensor to the 
vacuum feedthrough (item 4 in Fig.~\ref{fig:cryotip}), while a four wire setup is used outside 
of the vacuum system.   This arrangement, though not ideal, provides sufficiently 
low noise for the measurements.   An Agilent {\em E4980A} capacitance bridge able 
to excite the capacitor at 800~kHz while reading out at 20~samples$/$s is used for the 
more critical measurements shown below. This instrument, capable of resolving 0.1~fF at our sampling rate, 
is particularly useful when fast response is required in a PID loop to stabilize layer thicknesses. 
Alternative readouts schemes include an 
Andeen-Hagerling {\em 2550A} ultra-high precision capacitance bridge and a 
relatively inexpensive SRS {\em 715} bridge.     A home-built, custom tuned circuit 
mounted at the vacuum feedthrough was also used for some of the measurements.     
Conceivably, a similar circuit built on a chip and placed next to the sensor 
itself could provide the best noise performance for the system while simplifying the 
wiring.      In all cases two very stable 5~nF high voltage mica capacitors 
are used to decouple the readout electronics from the sensor. In this way, the sensor can be 
DC-biased up to a few kV, as it is required to attract Ba-ions in EXO. 
The temperature response of the sensors in vacuum was tested over the entire 15~K - 650~K range. 
The behavior during cooldown from 310~K to 15~K is shown in Fig.~\ref{vac}. 
This is found to be well modeled by accounting for the thermal expansion of the Al 
strips and the quartz substrate.   Overall the capacitance exhibits a small and positive 
temperature coefficient, as expected because the expansion coefficient of aluminum 
($k_{Al}$~=~23.1$\cdot10^{-6}$~K$^{-1}$) is substantially larger than that of quartz 
($k_{quartz}$~=~5.5$\cdot10^{-6}$~K$^{-1}$). This measurement 
is not appreciably affected by condensation because of the low base pressure 
of $< 10^{-8}$\,mbar, and the fast cooldown time of $\simeq$~30~s from about 
160~K to 20~K. A sensor calibration curve in vacuum is 
measured for every sensor type and used to 
correct all data presented below.

\section{Sensor Performance}\label{performance}

Once the offset capacitance of a particular sensor with its readout cables has been precisely 
established in the fluid phase (gas or liquid) of a dielectric material, its gain 
can be established by condensing a very thick layer ($>$~1000~nm) of the frozen phase, 
as determined by the saturation capacitance obtained in this way.  

\begin{figure}[!htb]
  \begin{center}
 \includegraphics[scale = 0.14]{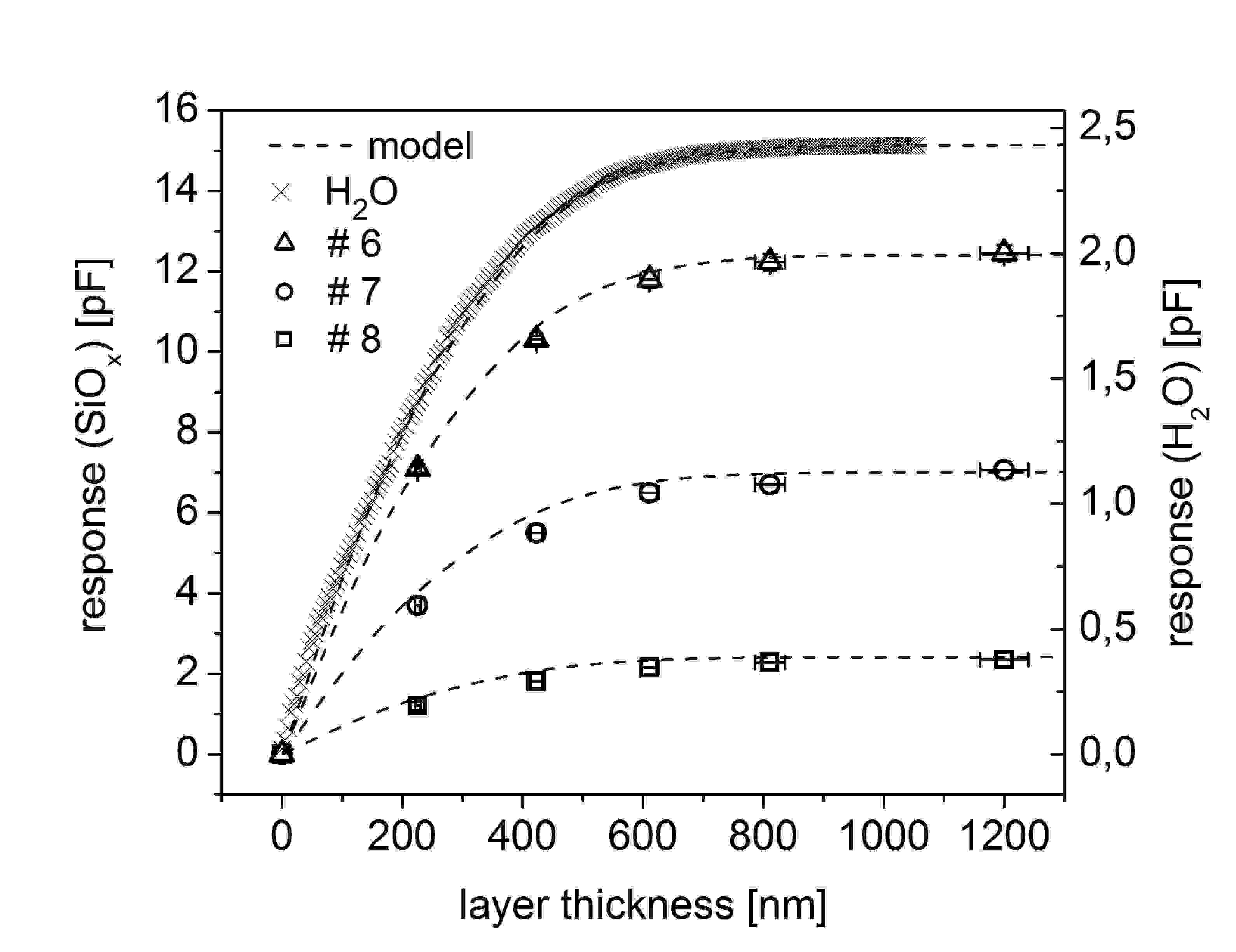}
  \caption{
The uppermost data set shows the capacitive response to the deposition of H$_2$O 
from vapor on a type 1 sensor and the simulated 
behavior (dashed line), referenced to the scale on the right.  The layer 
thickness is estimated using the calculated freezing rate and time, see text.   
The triangles, circles and rectangles denote 
the capacitive response of sensors of type 6, 7 and 8 respectively, to 
uniformly deposited silicon oxide layers. The lines represent the corresponding simulation.
The simulations are performed based on the detailed geometry of the sensor fingers, 
resulting in a very specific sensor response behavior from empty (no layer) 
to full (thick layers). 
The simulated curves are only scaled along the vertical axis to fit the data. }\label{fig:capcurve}
\end{center}
 \end{figure}

To verify that the saturation behavior is a proper normalization for the device at 
any thickness, a detailed calibration is performed for water and silicon oxide. These 
two calibratations involve very different systematic effects and their agreement with 
the finite element calculations demonstrates the accuracy of the devices. 
In the first case, water is frozen 
on a sensor by coupling the vacuum system to a small deionized water reservoir 
via a leak valve. The dielectric constant of water ice is $\epsilon_r$~$\sim$~3.2 at high 
frequencies \cite{ref:watereps}. The measurement is performed by establishing equilibrium 
between the leak rate of vapor into the vacuum system and pumping through a throttled  
turbo molecular pump. By measuring the difference in equilibrium pressure with the cryostat  
at 300~K (4.1$\cdot$10$^{-5}$\,mbar) and at 100~K (6.6$\cdot$10$^{-6}$\,mbar), 
a deposition rate of 8.4$\pm 0.8 \cdot 10^{14} $atoms/(s$\cdot$cm$^2$) can be 
calculated. This corresponds to a thickness growth rate of 0.27$\pm$0.03~nm/s, 
assuming crystalline ice formation. Figure~\ref{fig:capcurve} 
shows the measured capacitance for H$_2$O as a function of calculated layer thickness. 
In a second experiment, the response of the sensor was compared to a  
quartz crystal microbalance. For this test, uniform dielectric silicon oxide layers were 
repeatedly deposited on a wafer with sensors of types 6, 7 and 8 by means of low 
temperature vapor deposition in steps of about 200~nm. 
Capacitance was measured on different structures close to the center of the wafer and 
averaged. The dominant thickness uncertainty 
in this case originates from the wet-etch 
cleaning step before each further deposition step and is included in the error bars in the figure.

The dashed lines in Fig.~\ref{fig:capcurve} 
correspond to the result of a simulation of the response in each case 
using FEMLAB~\cite{ref:femlab}.  
The simulation is based on the detailed geometry of the aluminum fingers and is carried out for 
different dielectric layer thicknesses, resulting in a very specific sensor response 
curve. This curve is then vertically stretched to fit the measured saturation value without changing its  
characteristic shape. The good agreement between data and simulation shows that the 
physics of the device is well understood, once the exact value of the dielectric constant is 
extracted from data.

The saturation capacitance ($C_{sat}$) is also measured for several other materials 
in solid and liquid phase, as summarized in Tab.~\ref{tab:gasesdiel}.   
Using the same FEMLAB simulation, 
the sensitivity of the sensor is then calculated for the first $\simeq 100$~nm layer where 
the response is expected to be linear.    
This sensitvity parameter is also calculated from the saturation response of sensors 
fully submerged in liquids (although in practice, thin liquid layers are 
experimentally unattainable).     
The ratio between saturation capacitance and dielectric constant 
of the material is expected to be a property of the geometry of the sensor.   
This is verified, for two different sensor types, by the data shown in Tab.~\ref{tab:gasesdiel}.

\section{Accurate layer control with xenon}

\begin{figure}[!htb]
  \begin{center}
 \includegraphics[scale = 0.13]{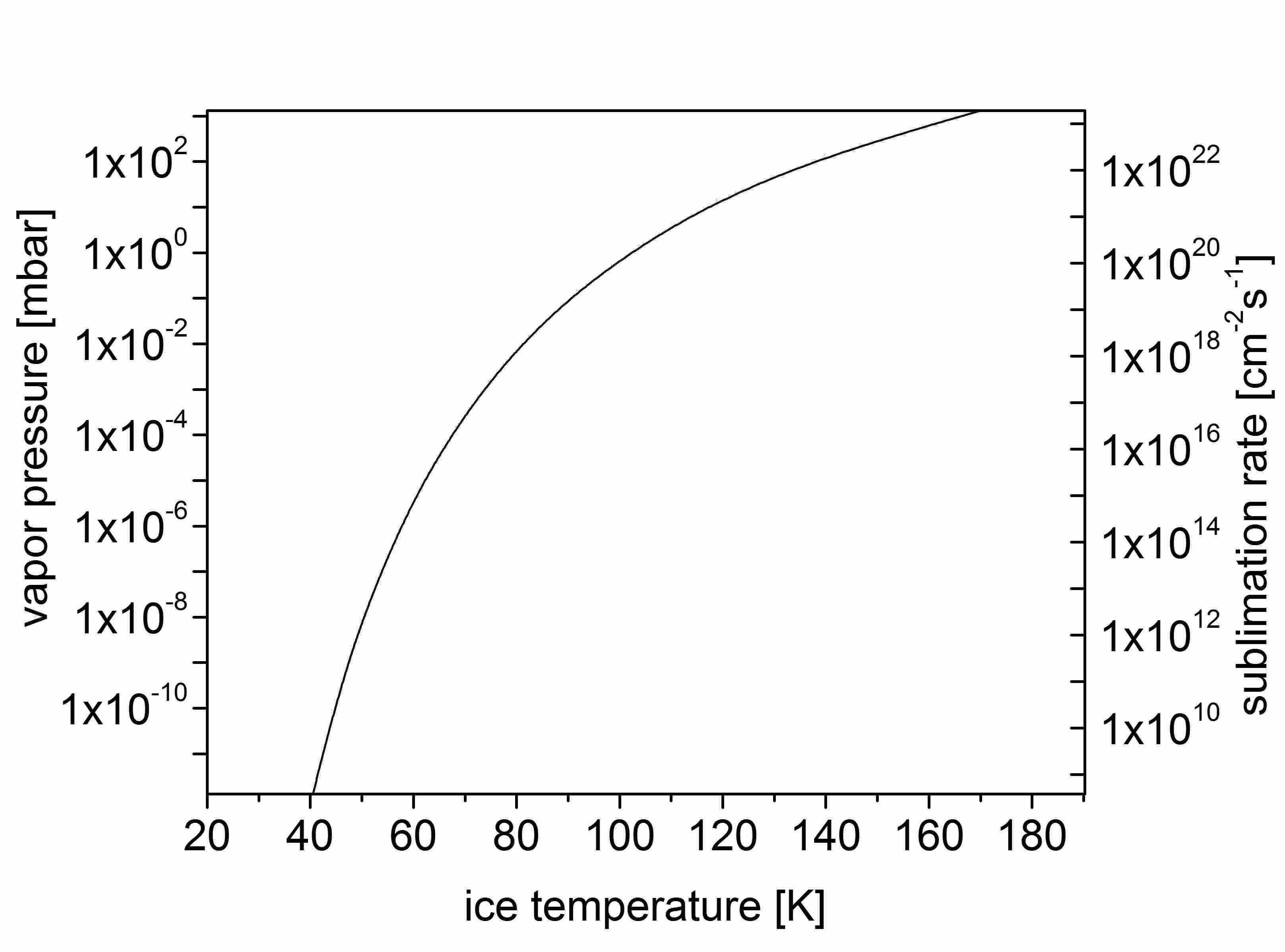}
  \caption{Calculated vapor pressure of solid Xe at different temperatures. On the right scale the 
   expected sublimation rate of Xe atoms based on the Hertz-Knudsen model in vacuum is shown. }\label{fig:hertzknudsen}
\end{center}
 \end{figure}

\begin{figure}[!htb]
  \begin{center}
 \includegraphics[scale = 0.13]{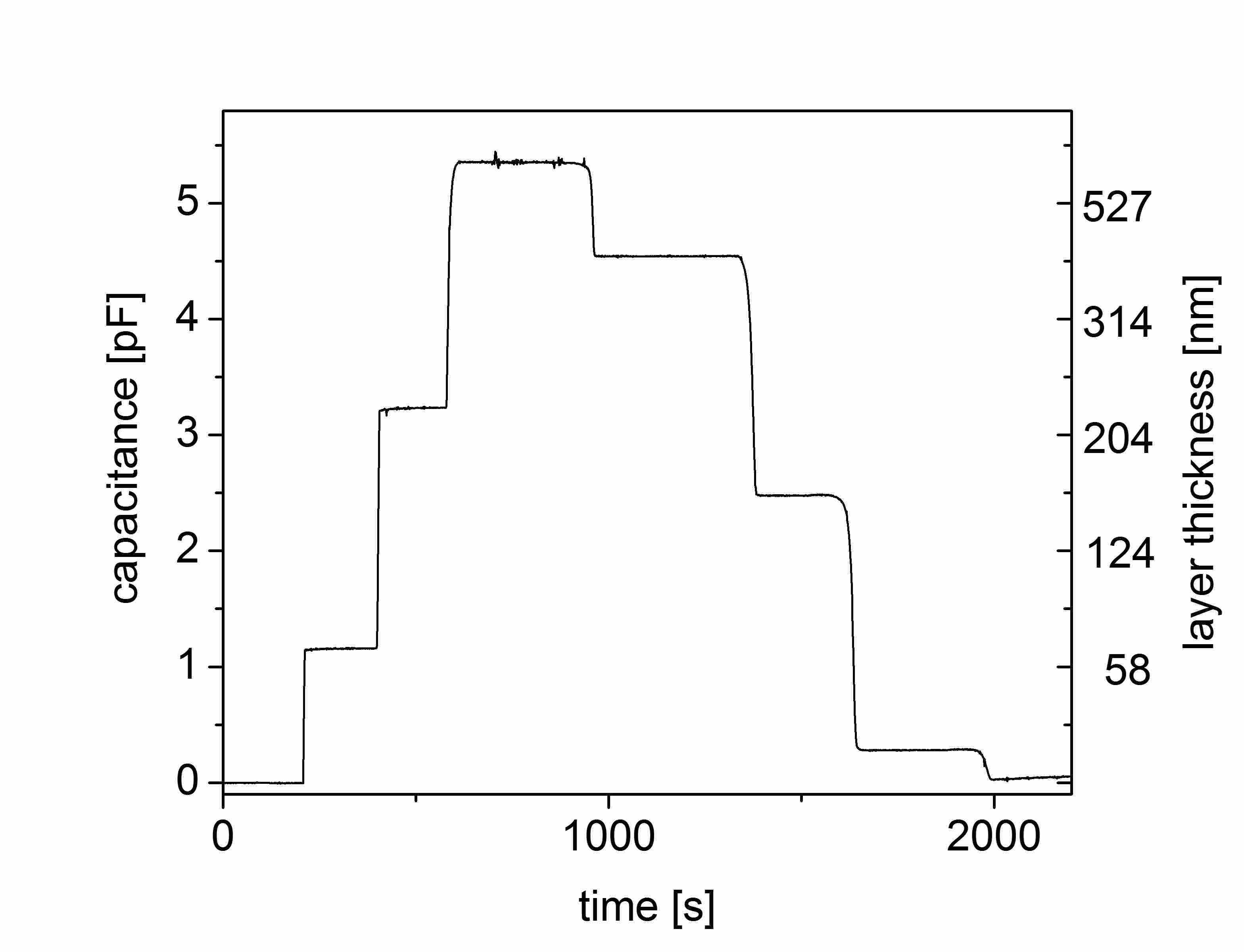}
  \caption{Capacitance of a type 5 sensor with different thicknesses of solid Xe
in $<$\,3$\cdot$10$^{-8}$\,mbar vacuum.  The solid Xe is deposited on the sensor, 
which is kept at $<$~30~K, by condensing a known amounts of gas.  
Sublimation is obtained by increasing the temperature of the sensor.   
The plateaus correspond to stable frozen Xe layers.}\label{fig:layerstability}
\end{center}
 \end{figure}

Several measurements were performed with the capacitive sensors to develop the ability 
to grow, stabilize and remove precisely controlled solid layers in different environments.  
The sublimation rate is described by the Hertz-Knudsen model \cite{hertzknudsen}, which 
approximates the net surface flux (in atoms cm$^{-2}$ s$^{-1}$) as
\begin{center}
\begin{equation}
\Phi = \alpha \frac{p_v - p}{\sqrt{2\pi mkT}}
\end{equation}
\end{center}
with $\alpha$ the sticking coefficient (typically 0.1-1) $p_v$ the equilibrium vapor 
pressure at the temperature of the surface $T$ and $p$ the ambient pressure (for the case of vacuum 
$p = 0$). The parameter $m$ is the mass of the atomic species sublimating and $k$ is Boltzmann's constant.
As an example, the values for $p_v$ for the case of Xe are computed using Ref.\,\cite{cite:shrim} and 
shown in Fig.~\ref{fig:hertzknudsen}. 
On the right axis the estimated 
sublimation rate in vacuum with $\alpha$ = 1 is indicated. 
In a first test, solid Xe layers are grown from the gas phase and 
kept stable for typical times of 100\,s in vacuum. To condense a layer, the tip with the sensor 
is cooled below 30~K and a leak valve (item 12 in Fig.~\ref{fig:cryotip}) is opened 
to set the partial pressure of Xe to $10^{-3}$\,mbar 
in the vacuum system. 
Controlled sublimation is obtained by increasing the probe tip temperature by gradually 
reducing the LHe flow through the cryostat. 
The data shown in Fig.~\ref{fig:layerstability} is collected with a sensor of type 5 
(see Tab.~\ref{tab:sensors}). 
The thicknesses of the layers in the figure 
are, from left to right, 67.7$\pm$0.1~nm, 224$\pm$0.2~nm, 750~nm (thicker than saturation), 403.1$\pm$0.1, 
159.9$\pm$0.5~nm and 15.8$\pm$0.1~nm. No sublimation within the sensitivity of the device is 
observed in a $<$\,3$\cdot$10$^{-8}$\,mbar vacuum at a tip temperature of $\simeq$~30~K. 
The errors on these measurements are dominated by the noise of the readout cables and the 
SRS\,715 capacity bridge, which was used for this test and set to "slow" readout (10\,kHz 
excitation and 9 sample averaging at 4 samples/s using a two wire measurement). 

\begin{figure}[!htb]
  \begin{center}
 \includegraphics[scale = 0.13]{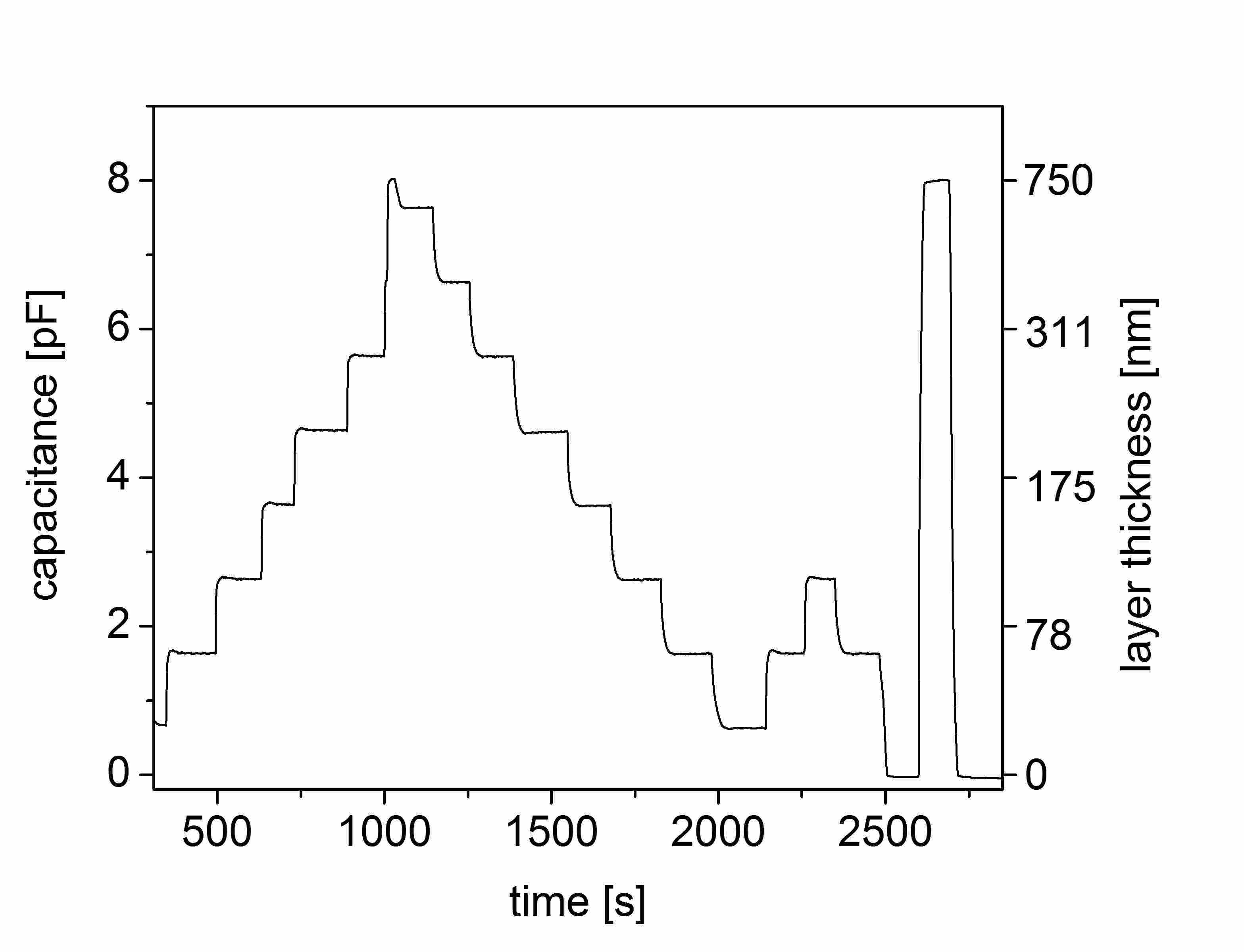}
  \caption{Dynamically stabilized solid Xe layers grown and sublimated at constant pressure 
of 700\,mbar Xe gas, 
using a sensor type 9 and active feedback on the capacitance of the sensor. 
The left and right scales are, respectively, the sensor capacitance and thickness. 
At 7.98\,pF the sensor response 
saturates.  Stable layers are, again, obtained for periods of many minutes.
}\label{fig:xegas}
\end{center}
 \end{figure}

A somewhat more challenging task consists of actively growing and stabilizing 
a solid Xe layer condensed from an infinite 
supply of Xe gas at 700\,mbar and a temperature of $\sim$ 155\,K.  
In this case the temperature of the probe has to be actively maintained in a feedback 
loop using the capacitance readout from the sensor.    
The LHe flow is kept constant while some heat is added to the tip by 
powering the tungsten wire (item 4 in Fig.~\ref{fig:cryotip}) from a PID control 
loop realized within the LabView program. Stabilization is based directly 
on the layer thickness, which is calculated online from the capacitance of the sensor, 
correcting for the temperature coefficient, as explained. In this case, the capacitance is measured 
using the Agilent bridge with "fast" readout setting (20~samples/s) and 
4 sample averaging at 800~kHz excitation frequency. The PID loop has a cycle length of 
50~ms. In Fig.~\ref{fig:xegas} 
a series of PID-stabilized layers based on a sensor of type 9 is shown. 
The sensor has a saturated capacitance for SXe at 155~K of 7.979$\pm$0.003~pF. This 
corresponds to a response of 16.62~fF/nm for layers $<$~100~nm (see Tab.~\ref{tab:gasesdiel}).
The layer thicknesses chosen for this test 
are 24.3$\pm$0.7\,nm, 62.1$\pm$1.3\,nm, 107.2$\pm$1.5\,nm, 156.7$\pm$2.3\,nm, \
214.5 $\pm$2.8\,nm, 284.4$\pm$5.1\,nm, 
383$\pm$10\,nm and \ 567 $\pm$40\,nm, where the errors are the maximum deviation 
from the average value during the PID stabilization. The errors are dominated by the 
stability of the PID control loop.   
Again, layers stable for several minutes are clearly achieved. 
With the same setup the temperature of the cryostat was changed from 155\,K to 10\,K and the  
expected increase in the dielectric constant of 6.4\% with the change in density of the solid 
\cite{ref:sears} was indeed observed.

\begin{figure}[!htb]
  \begin{center}
 \includegraphics[scale = 0.13]{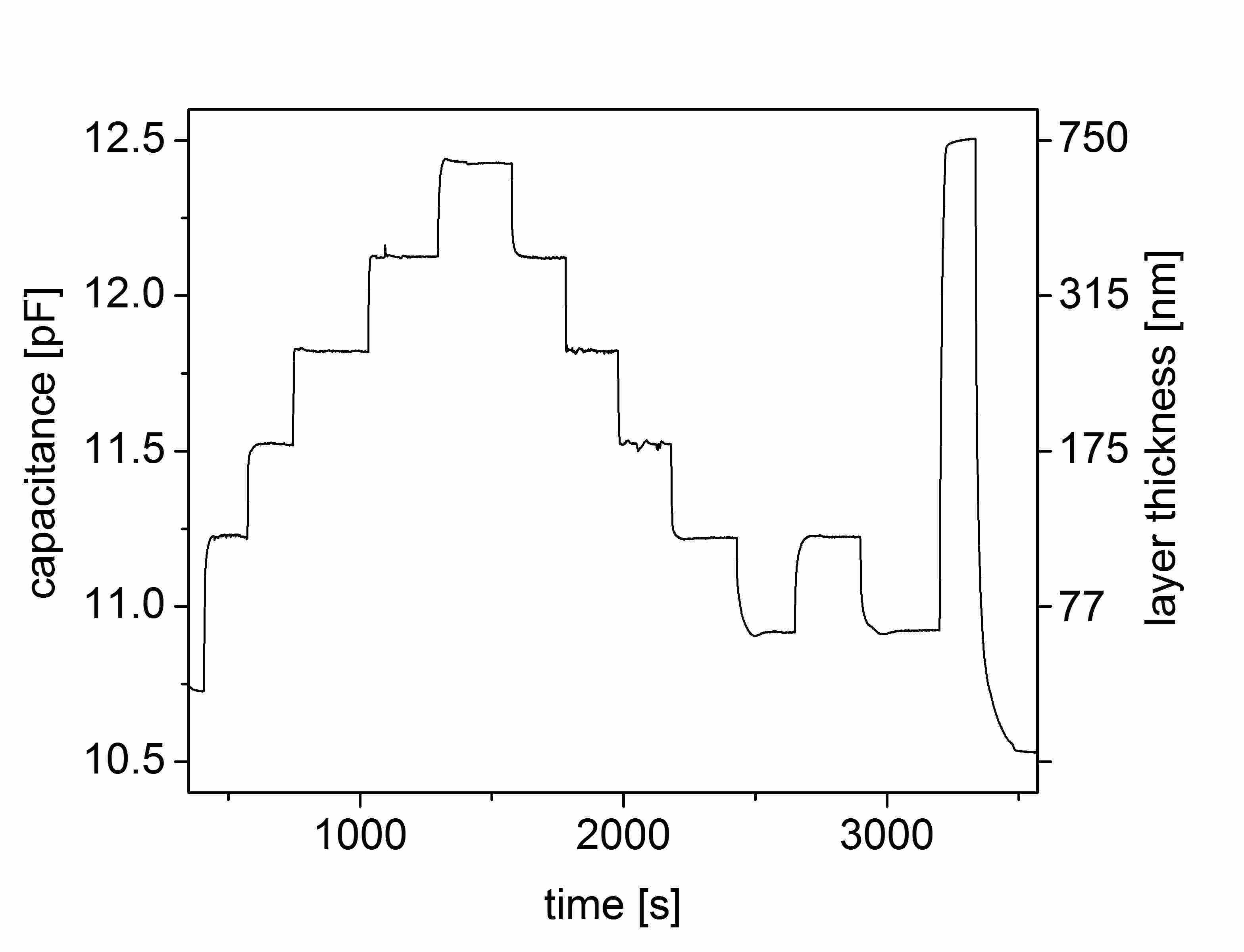}
  \caption{Solid frozen from LXe, using a sensor type 10 and active feedback. 
At 12.5\,pF the sensor saturates. The offset of 10.52\,pF is due to the dielectric constant of LXe. 
}\label{fig:xeliqu}
\end{center}
 \end{figure}

In order to freeze Xe from the liquid phase, the gate valve (item 14 in Fig.~\ref{fig:cryotip}) 
below the probe tip is opened and the bellows (item 4 in Fig.~\ref{fig:cryotip}) is compressed, 
so that the tip assembly with a sensor of type 10 reaches into a cold cell filled with liquid Xe 
(item 15 in Fig.~\ref{fig:cryotip}). A customized LabView PID loop is used 
to control the current through the tungsten heater. Data were taken with the Agilent capacity bridge 
also using the "fast" readout setting and only two samples averaging, which allows for a shorter PID cycle length 
of 20~ms. In order to keep conditions stable, the temperature of the liquid Xe cell is stabilized with 
a separate PID control loop to 163.00$\pm$0.03\,K, which contains an unknown offset due to the position 
of the thermocouple on the outside of the LXe copper cell. The data for this case are shown in 
Fig.~\ref{fig:xeliqu}. 
As in the previous case the noise of the measurement is dominated by the PID 
control loop. Because of the smaller 
difference in dielectric constant between liquid and solid phases, 
similar errors on the capacitance result in larger uncertainties in the layer thickness. 
The layer thickness, as in the previous tests, 
is calibrated using the FEMLAB simulation, with the dielectric constants 
appropriate for the LXe environment. The steps in the graph correspond to layer thicknesses 
of 32.9$\pm$0.4\,nm, 119.7$\pm$0.8\,nm, 185$\pm$3\,nm, 262$\pm$2\,nm, and 383$\pm$8\,nm,  
where the error denotes the maximum deviation from the average value over the time the layer is stabilized.  
The last peak refers to the saturation capacitance of 12.48\,pF for solid Xe, while the value of 10.51\,pF 
for pure liquid is measured at the end of the plot. The corresponding sensitivity for $<$~100~nm layers 
is 38.75~fF/nm for SXe in Xe gas and 6.14~fF/nm for SXe with the sensor immersed in LXe.  
This measurement reproduces the expected behavior of LXe and SXe with dielectric constants 
$\epsilon_{r,LXe}$\,=\,1.88 (161.35\,K, $\rho$ =  2.98\,g/cm$^3$)~\cite{ref:dielconst} and 
$\epsilon_{r,SXe}$\,=\,2.2 (160\,K, $\rho$ = 3.54\,g/cm$^3$)~\cite{ref:sxedensity}.

\begin{figure}[!tb]
  \begin{center}
 \includegraphics[scale = 0.068]{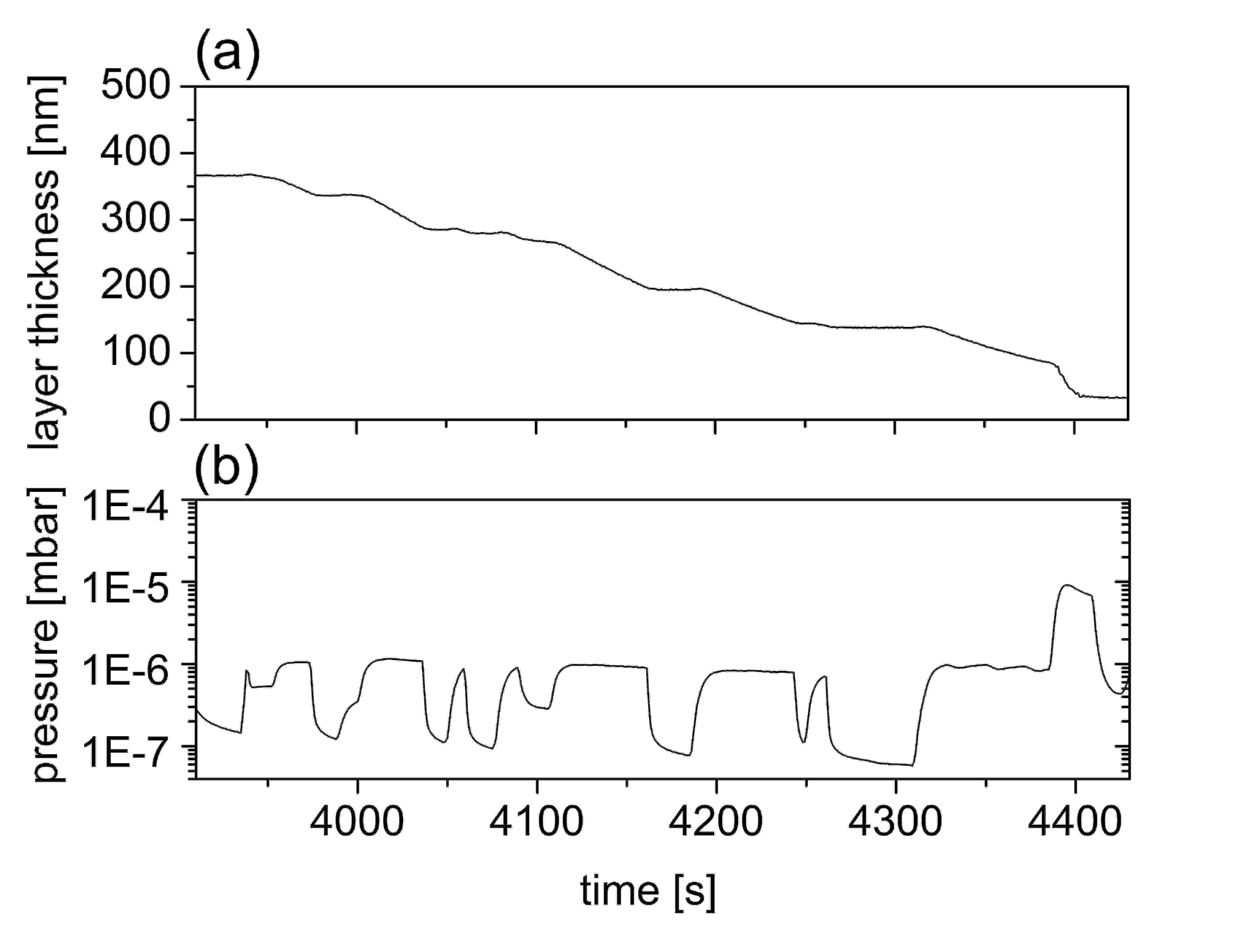}
  \caption{Controlled sublimation of SXe layers. Panel (a) shows the layer thickness 
as a function of time. The pressure in the vacuum chamber during controlled 
sublimation is shown in panel (b). }\label{fig:evap}
\end{center}
 \end{figure}

In a final test, layers are sublimated in a controlled way to keep the partial pressure 
of the evaporated gas at a low level. This is important for the EXO experiment, 
where strict bounds exist on the amount of Xe tolerable in the ion trap vacuum system \cite{jessePRL}. 
The tungsten heater on the back side of the sensor is used to sublimate the layers, while the 
cryostat is kept cold. The correlation of sublimation rate and pressure in the 
vacuum system is shown in Fig.\,\ref{fig:evap}. 
Here, the vacuum chamber with a volume of 
1.5~l is pumped with a 70\,l/s turbo molecular pump. The plot indicates that an evaporation rate 
of 2\,nm/s results in a partial pressure of 
1$\cdot$10$^{-6}$\,mbar. Scaled to the EXO ion trap system with a pumping speed of 500\,l/s and a volume 
of 20\,l, this corresponds to a partial pressure of 1$\cdot$10$^{-8}$\,mbar, 
well below the acceptable limit of about 5$\cdot$10$^{-6}$\,mbar.

\section{Conclusion}\label{conclusion}

We have developed a versatile and simple microfabricated sensor for accurate capacitive 
measurements of thin layers of dielectric substances. The 
sensor can be calibrated for many substances due to its asymptotic saturation 
behavior for thick deposited layers. This asymptotic behavior is in good agreement with an 
electrostatic field simulation and the relative dielectric constant values from the literature. 
Data was presented for the specific 
application that led to this development, the controlled growth, stabilization and 
removal of solid Xe layers in gaseous Xe, liquid and vacuum environments. 
The sensors presented, with appropriate readout schemes, are capable 
of controlling solid Xe layers in Xe gas and liquid with sensitivities of 38.75~fF$/$nm and 6.2~fF$/$nm, 
respectively. Other applications of the sensor include 
the control of coatings with different dielectric materials from gaseous or liquid 
environments and the characterization of dielectric emulsions in situations 
where ruggedness and miniaturization are important.

\section{Acknowledgements}

The essential support from the staff at the Stanford Nanofabrication Facility is gratefully 
acknowledged.     We also thank B.~Young for advice and technical help and D.~Osheroff 
for the loan of the Andeen-Hagerling capacitance bridge.     Finally we thank the Agilent 
Corporation for the loan of one of their {\em E4980A} capacitance bridges. 

\onecolumn

\onecolumn

\clearpage
 
\renewcommand{\arraystretch}{1}
\begin{table}[!htb]
\begin{center}
\begin{tabular}{l c c c c c}
  {\bf Sensor}   &  {\bf Spacing $s$} & {\bf Thickness $t$ }&  {\bf Width $w$ } &   {\bf Area}   & {\bf Capacitance }   \\
   {\bf }   & {\bf [$\mu$m]} & {\bf [$\mu$m]} & {\bf [$\mu$m]}& {\bf [mm$^2$]} & {\bf [pF]   }       \\
\hline \hline
  1   &          1.50 &            1.85 &       0.8  &      0.69  &        7.3    \\
  2   &          1.50 &            1.85 &       2.5  &      0.69  &         4.1    \\
  3   &          1.50 &            1.85 &       5.5  &      0.69  &         3.2    \\
  4   &          1.50 &            1.85 &       11.5 &      0.69  &         2.3     \\
  5   &          1.50 &            1.85 &       0.8  &      2.84  &         23.1    \\
  6   &          1.50 &            1.85 &       2.5  &      2.84  &         14.0    \\
  7   &          1.50 &            1.85 &       5.5  &      2.84  &         11.6   \\
  8   &          1.50 &            1.85 &       11.5 &      2.84  &         9.1    \\
  9   &          1.47 &            2.00 &       2.5  &      2.26  &         23.0    \\
 10   &          1.47 &            2.00 &       1.5  &      2.26  &         28.0    \\
%
\hline \\
\end{tabular}
\medskip
\caption{Properties of all sensors produced.   Devices in the first part of the list are
tested in detail and described in this paper.      Items 9 and 10 have larger connection pads. 
}
\label{tab:sensors}
\end{center}
\end{table}
\renewcommand{\arraystretch}{1}

\renewcommand{\baselinestretch}{1}

\newpage
\clearpage

\renewcommand{\baselinestretch}{1}

\renewcommand{\arraystretch}{1}
\begin{table}[!htb]
\begin{center}
\begin{tabular}{l r r c r r r}
\multicolumn{1}{l} {\bf Substance} & 
\multicolumn{1}{c} {\bf C$_{sat}$ } &  
\multicolumn{1}{c} {\bf Sensitivity}  &
\multicolumn{1}{c} {\bf Sensor} & 
\multicolumn{1}{c} {\bf $\epsilon_r$} & 
\multicolumn{1}{c} {\bf C$_{sat}$/$\epsilon_r$} \\

\multicolumn{1}{l} {\bf } & 
\multicolumn{1}{c} {\bf [pF]} &  
\multicolumn{1}{c} {\bf [fF/nm]}  &
\multicolumn{1}{c} {\bf } & 
\multicolumn{1}{c} {\bf } & 
\multicolumn{1}{c} {\bf [pF]} \\

\hline \hline
     H$_2$O (solid) &    2.40$\pm$0.02       &   7.4$\pm$0.1     &       1          &           $\sim$~3.2\cite{ref:watereps}  &     0.75$\pm$0.01            \\
     Xe (liquid)    &    1.260$\pm$0.005    &   3.9$\pm$0.1     &       1          &            1.880 \cite{ref:dielconst}        &      0.67$\pm$0.01       \\
     Xe (solid, 100\,K) & 1.515$\pm$0.005   &   4.7$\pm$0.1     &       1          &            2.24 \cite{ref:dielconst}         &      0.67$\pm$0.01        \\
     CO$_2$ (solid) &    1.43$\pm$0.01      &   4.4$\pm$0.1     &       1          &            2.12 \cite{ref:dielconstco2}           & 0.68$\pm$0.01        \\
     Xe (solid, 30\,K) & 5.241$\pm$0.005    &   16.3$\pm$0.1    &       6          &            2.25 \cite{ref:hilt,ref:sears} &      2.33$\pm$0.01           \\
     Kr (solid, 30\,K) &    4.118$\pm$0.005 &   12.8$\pm$0.1    &       6          &            1.66 \cite{ref:dielconst}         &   2.48$\pm$0.01            \\
     Ar (liquid, 87\,K)     &    3.61$\pm$0.05 & 11.2$\pm$0.2   &       6          &            1.50  \cite{ref:dielconst} &      2.41$\pm$0.03           \\
     N$_2$ (liquid, 70\,K) &    3.55$\pm$0.05  & 11.0$\pm$0.2   &       6          &            1.45  \cite{ref:dielconst} &     2.44$\pm$0.04        \\
\hline
     Xe (solid, 15\,K) & 5.36$\pm$0.004     &   11.16$\pm$0.01  &       5          &            2.25 \cite{ref:hilt,ref:sears}   &      2.38$\pm$0.01           \\
     Xe (solid, $\simeq$~155~K) & 7.979$\pm$0.003 & 16.62$\pm$0.01 &    9          &            2.20 \cite{ref:hilt,ref:sears}    &     3.59$\pm$0.01          \\
     Xe (liquid) & 10.512$\pm$0.002         &   21.89$\pm$0.01         & 10        &            1.880 \cite{ref:dielconst}       &      5.59$\pm$0.02          \\
     Xe (solid, $\simeq$~155~K) & 12.484$\pm$0.002  &   38.75$\pm$0.01 & 10        &            2.20 \cite{ref:hilt,ref:sears}   &      5.62$\pm$0.02          \\


\hline \\
\end{tabular}
\caption{Sensor parameters for different phases of various media.  
The saturation capacitance, $C_{sat}$ is given in the second column. The sensitivity, 
in fF/nm, is derived from the measured $C_{sat}$ and the saturation thickness 
calculated with FEMLAB.   In general, this sensitivity only applies to the linear 
region where the thickness is $<$~100~nm.   The sensor type is given in the fourth 
column, with reference to Tab.~1.   The fifth and sixth columns show, respectively, 
the relative dielectric constant $\epsilon_r$ (from the literature) and the ratio between 
columns 2 and 5. This last parameter is a function of the sensor geometry only 
and is expected to be constant for a given sensor type, as it is generally found.
%
All literature values of $\epsilon_r$ 
are used without error. The parameters for media in liquid phase are relevant 
here because their value of $C_{sat}$ has to be subtracted to obtain the 
behavior of a sensor on which solid is grown from liquid.}\label{tab:gasesdiel}
\end{center}
\end{table}
\renewcommand{\arraystretch}{1}

\end{document}